\documentclass[prl,aps,showpacs,twocolumn,superscriptaddress,notitlepage]{revtex4}
\usepackage{amsmath}
\usepackage{graphicx}
\usepackage{bm}
\usepackage{amsbsy}
\usepackage{amsfonts}
\usepackage{amsthm}
\usepackage[dvips]{color}

\def\mn#1{\langle #1 \rangle}

\newcommand{\cit}[1]{#1}

\newcommand{\ket}[1]{|#1\rangle}

\newcommand{\CG}{coarse-grain}

\newcommand{\PDF}{probability distribution}
\newcommand{\PC}{phase covariant}
\newcommand{\MP}{measure-and-prepare}

\newcommand{\intphi}{\phi}

\newcommand{\BS}{Bloch Sphere}

\newcommand{\Hm}{H}

\begin{document}

\title{Coarse-graining makes it hard to see micro-macro entanglement}

\author{Sadegh Raeisi}
\affiliation{Institute for Quantum Information Science, University of Calgary, Alberta T2N 1N4, Canada}
\affiliation{Institute for Quantum Computing, University of Waterloo, Ontario N2L 3G1, Canada}
\author{Pavel Sekatski }
\affiliation{Group of Applied Physics, University of Geneva, Geneva, Switzerland}
\author{Christoph Simon}
\affiliation{Institute for Quantum Information Science, University of Calgary, Alberta T2N 1N4, Canada}

\begin{abstract}
Observing quantum effects such as superpositions and entanglement in macroscopic systems requires not only a system that is well protected against environmental decoherence, but also sufficient measurement precision. Motivated by recent experiments, we study the effects of coarse-graining in photon number measurements on the observability of micro-macro entanglement that is created by greatly amplifying one photon from an entangled pair. We compare the results obtained for a unitary quantum cloner, which generates micro-macro entanglement, and for a measure-and-prepare cloner, which produces a separable micro-macro state. We show that the distance between the probability distributions of results for the two cloners approaches zero for a fixed moderate amount of coarse-graining. Proving the presence of micro-macro entanglement therefore becomes progressively harder as the system size increases.
\end{abstract}

\maketitle

How does the classical world emerge from quantum physics? It is now widely recognized that decoherence is one important factor. As the size of physical systems increases, it becomes harder to isolate them completely from their environment, and the interaction with the environment destroys quantum features such as superpositions and entanglement \cite{Zurek-CtoQ-03,Zurek-CtoQ-91}. However, this is not the only reason why quantum effects are difficult to detect at the macroscopic scale. Measurement precision also seems to be essential. For example, in 1979 Mermin \cite{mermin} studied a singlet state of two large spins $J$. He showed that this state can violate a Bell inequality for arbitrarily large $J$, thus proving entanglement and quantum non-locality, but the necessary angular resolution of the measurements decreases as $\frac{1}{J}$, making them harder and harder to perform for increasing $J$. Later Peres \cite{peres} showed for the same state that its spin correlations can be reproduced by a classical model if the resolution in the measurement of the spin projections $M$ is much worse than $\sqrt{J}$. Conversely, it was shown in Ref. \cite{entlaser} for a closely analogous multi-photon state that macroscopic entanglement can be proved by measuring spin (or, more precisely, Stokes parameter) correlations, provided that the precision of the photon counters is better than $\sqrt{N}$, where $N = 2J$ is the total number of photons.

The mentioned results concerned the entanglement of two macroscopic systems (spins), and they were largely theoretical, although Ref. \cite{eisenberg} succeeded in demonstrating entanglement for up to 12 photons for states of the form considered in Ref. \cite{entlaser}. More recently Ref. \cite{demartini} claimed the experimental creation and detection of ``micro-macro'' entanglement between one photon in one spatial mode and of order $10^4$ photons in another spatial mode, by greatly amplifying one photon belonging to an initial entangled pair. In their argument, the authors of Ref. \cite{demartini} used very coarse-grained (binary) measurements of their multi-photon state, in combination with an entanglement criterion that had been derived for individual photons (qubits) in Ref. \cite{eisenberg}. It was subsequently shown in Refs. \cite{sekatskiprl,sekatskipra} that this criterion is not conclusive in the multi-photon case because it can also be violated with separable states. In particular it was shown in Ref. \cite{sekatskipra} that for extremely coarse-grained measurements of the type considered in Refs. \cite{demartini,sekatskiprl} a ``measure-and-prepare'' type amplification strategy, which destroys all entanglement, yields results that are indistinguishable from those obtained by a unitary quantum cloner, which creates micro-macro entanglement in the ideal case.

Demonstrating entanglement in this system with very coarse-grained measurements is thus impossible, unless supplementary assumptions are made \cite{dmassumptions}. On the other hand, it is clear that under close-to-ideal conditions micro-macro entanglement analogous to Schr\"{o}dinger's famous cat example \cite{cat} would indeed be created in the system of Ref. \cite{demartini}. The spin correlation criterion of Ref. \cite{entlaser} can be adapted to the micro-macro situation. Entanglement could be proved experimentally using this adapted criterion provided that there is not too much loss, and, most importantly, that the photon counters can count large photon numbers with an accuracy at the single-photon level \cite{sekatskiprl,sekatskipra}. However, this becomes very difficult for large photon numbers. There is thus strong experimental and theoretical motivation to study the effects of coarse-graining on the observability of micro-macro entanglement in this system more generally. Is the requirement for single-photon level resolution just an unfortunate feature of the particular criterion used in
Refs. \cite{sekatskiprl,sekatskipra}, or is it more fundamental? This is the question that we study in the present paper.

\begin{figure*}[t]
\centering{}\includegraphics[width=\textwidth]{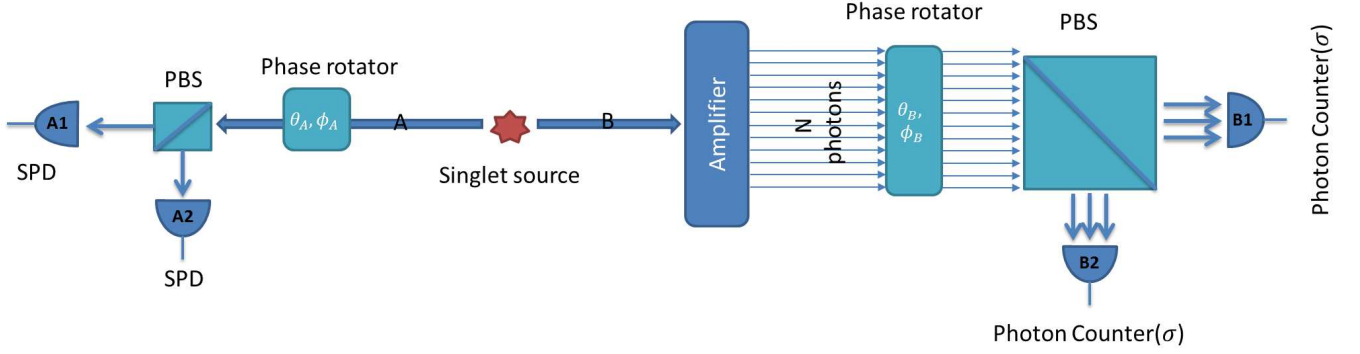}\caption{Schematic of the experiments considered in the present paper. A source produces an entangled photon pair in a polarization singlet state. Photon A is detected in an arbitrary polarization basis with the help of the phase rotator, polarizing beam splitter (PBS) and single-photon detectors (SPD). Photon B is amplified, either by a unitary cloner, which preserves the initial entanglement, or by a measure-and-prepare cloner, which completely destroys the entanglement. The resulting multi-photon state is detected in a similar way, where the single-photon detectors are now replaced by photon counters which detect the number of photons in each polarization mode. The coarse-graining parameter $\sigma$ characterizes the precision with which the photon number can be measured.}
\label{fig:process}%
\end{figure*}

Figure 1 is a schematic representation of the system that we are considering \cite{demartini,sekatskiprl}. A source produces pairs of entangled photons in two distinct spatial modes A and B in the polarization singlet state, $|\psi_-\rangle=\frac{1}{\sqrt{2}}(a^{\dagger}_h b^{\dagger}_v-a^{\dagger}_v b^{\dagger}_h)|0\rangle$, where $|0\rangle$ is the vacuum for all modes.
Identifying the $h$ and $v$ polarization with the north and south poles of the Bloch sphere, one can introduce modes for general polarization directions characterized by spherical angles $\theta$ and $\phi$ by the relation $a_{\theta,\phi}=\cos \frac{\theta}{2} e^{i\frac{\phi}{2}} a_h + \sin \frac{\theta}{2} e^{-i \frac{\phi}{2}} a_v$. Equatorial polarization modes correspond to $\theta=\pi/2$. Different choices of the phase $\phi$ give different equatorial bases (e.g. left and right circular polarization, or two orthogonal diagonal polarizations), and we will sometimes use the simplified notation $a_{\phi}\equiv a_{\pi/2,\phi}$ and $a_{\phi \perp} \equiv a_{\pi/2,\phi+\pi}$ for the two orthogonal modes corresponding to the basis defined by $\phi$. The singlet state keeps its form in any basis, in particular $|\psi_-\rangle=\frac{1}{\sqrt{2}}(a^{\dagger}_{\phi} b^{\dagger}_{\phi\perp}-a^{\dagger}_{\phi\perp} b^{\dagger}_{\phi})|0\rangle=\frac{1}{\sqrt{2}}(|\phi\rangle_A |\phi_{\perp}\rangle_B-|\phi_{\perp}\rangle_A |\phi\rangle_B)$, where we have introduced the notation $|\phi\rangle_A \equiv a^{\dagger}_{\phi}|0\rangle$ for a single $\phi$-polarized photon in mode A etc.

The photon in mode A is measured directly. The photon in mode B is amplified. We first consider the case where this amplification is done by a unitary phase-covariant quantum cloner \cite{brussPCC}. This type of cloner makes good copies only of input states that lie on the equator of the Bloch sphere. It can be realized based on stimulated parametric down-conversion \cite{nagali}. The Hamiltonian describing this process is
\begin{equation}
\Hm=i\chi b_{h}^{\dagger}b_{v}^{\dagger}+h.c.=i \frac{\chi}{2}(b^{\dagger 2}_\phi+b^{\dagger 2}_{\phi \perp})+h.c.
\label{eq:PC_Hamiltonian}
\end{equation}
It has the same form for any choice of equatorial basis, which is why the cloning process is phase covariant. Applying the unitary cloning operation $U=e^{-itH}$ to the photon in mode B results in the state
\begin{equation}
U|\psi_-\rangle=\frac{1}{\sqrt{2}} \left( |\phi\rangle_A |\Phi^{\phi \perp}\rangle_B-
|\phi_{\perp}\rangle_A |\Phi^{\phi}\rangle_B  \right),
\label{outputstate}
\end{equation}
with the multi-photon states
\begin{eqnarray}
\left| \Phi ^{\phi }\right\rangle =\sum\limits_{k,l=0}^{\infty
}\gamma _{kl}\frac{\sqrt{(2k+1)!(2l)!}}{k!l!}|2k+1\rangle_\phi |2l\rangle_{\phi\perp} \\
\left| \Phi ^{\phi \perp }\right\rangle =\sum\limits_{k,l=0}^{\infty
}\gamma _{kl}\frac{\sqrt{(2k+1)!(2l)!}}{k!l!}| 2k\rangle_\phi |2l+1\rangle_{\phi\perp}
\end{eqnarray}
with $\gamma _{kl}\equiv C^{-2}(-\frac{\Gamma }{2})^{k}\frac{\Gamma }{2}%
^{l}$, $C\equiv \cosh g$, $\Gamma \equiv \tanh g$, where $g=\chi t$ is the gain of the amplifier, and $|2k\rangle_{\phi}$ is the Fock state with $2k$ photons in mode $b_{\phi}$ etc. Due to the structure of $H$, $\ket{\Phi ^{\phi }}$ only contains terms with odd numbers of $\phi$-polarized photons and even numbers of $\phi_{\perp}$-polarized photons, whereas the opposite holds for $\ket{\Phi ^{\phi \perp}}$. The two macro-states $\ket{\Phi ^{\phi }}$ and $\ket{\Phi ^{\phi \perp}}$
are thus orthogonal to each other, and the micro-macro state of Eq. (\ref{outputstate}) is maximally entangled, which is consistent with the fact that the phase-covariant cloning transformation $U$ is unitary. For more details see \cit{\cite{demartini,sekatskiprl,sekatskipra,dmassumptions}}.

In experiments one aims to infer the existence of entanglement from the results of measurements, ideally without making assumptions about the process that led to those results. In this paper we will consider photon counting measurements, which are formally equivalent to spin measurements. For example, one can define the Stokes parameters
$J_{z}=b^{\dagger}_h b_h-b^{\dagger}_v b_v$, $J_x=b^{\dagger}_{\pi/2,0} b_{\pi/2,0}-b^{\dagger}_{\pi/2,\pi} b_{\pi/2,\pi}$, $J_y=b^{\dagger}_{\pi/2,\pi/2} b_{\pi/2,\pi/2}-b^{\dagger}_{\pi/2,3\pi/2} b_{\pi/2,3\pi/2}$ for system B, with $\sigma_z=|h\rangle \langle h|-|v\rangle \langle v|$ etc. the corresponding single-photon observables for system A. One can also define analogous observables for general directions (other than x, y, z).

Based on the results of Ref. \cite{entlaser}, one can show \cite{sekatskiprl} that all separable states fulfill
$|\mn{\vec{\sigma}_A\cdot \vec{J}_B}| \leq
\mn{N_{B}}$, i.e. the spin-spin correlation between the micro and macro systems is bounded by the total number of photons in the macro system. In contrast, the state of Eq. (\ref{outputstate}) gives
$|\mn{\vec{\sigma}_A\cdot \vec{J}_B}| -
\mn{N_{B}}=2$. Both $\vec{J}_B$ and $N_B$ involve measurements of large photon numbers. One sees that if these measurements are inaccurate by just a few photons, the presence of entanglement cannot be proven using this criterion. This leads to the question if the requirement for single-photon level resolution is a feature just of this particular criterion, or if it is in fact more general.

Here we address this question by comparing the probability distributions of results under coarse-graining for the (entanglement-preserving) unitary phase-covariant cloner described above, and for (entanglement-breaking) measure-and-prepare phase-covariant cloners \cite{sekatskipra,pomarico} described below. For comparing the two cases, we consider general single-photon polarization measurements characterized by angles $\theta_A$ and $\phi_A$ on A and photon counting measurements in arbitrary polarization bases $\theta_B, \phi_B$ on B. Without loss of generality, one can adopt the point of view that the measurement on A projects the photon in B before amplification into a well-defined state $|\theta_A,\phi_A\rangle$ (because the two initial photons are prepared in a singlet state). The measurement in A thus defines the input state of the amplifier in B. As the amplification processes are \PC, only the difference $\phi_{A}-\phi_B \equiv \Delta \phi$ is important.

The measure-and-prepare cloners are based on measuring the single photon in mode B in a random equatorial polarization basis, and then preparing a multi-photon state whose form depends on the measurement result. This procedure clearly destroys the entanglement between A and B, since a measurement is performed. The output state of a measure-and-prepare cloner is
\begin{align}
\rho_{mp}=\frac{1}{\pi}\int d\intphi(P^{+}(\theta_A,\phi_A , \intphi) & \mid\Psi_{\intphi}\rangle\langle\Psi_{\intphi}\mid\nonumber \\
+P^{-}(\theta_A,\phi_A , \intphi) & \mid\Psi_{\intphi\perp}\rangle\langle\Psi_{\intphi\perp}\mid),\label{eq:MP-macrostate}\end{align}
where $P^{+}(\theta_A,\phi_A ,\intphi)=|\langle \theta_A,\phi_A|\intphi\rangle|^2$ and $P^{-}(\theta_A,\phi_A, \intphi)=|\langle\theta_A,\phi_A|\intphi_{\perp}\rangle|^2$ indicates the probability of getting $\pm$
outcome for the measurements in the equatorial basis characterized by $\phi$, and the state $\mid\Psi_{\intphi (\perp)}\rangle$
is the multi-photon state that the cloner generates if the
measurement outcome is $|\intphi_{(\perp)}\rangle$.

The state that is prepared depends on the specific measure-and-prepare cloner considered. For simplicity we will compare the unitary and measure-and-prepare cloners for a fixed total number of output photons $N$. Since we are considering only photon counting measurements (which project onto subspaces of fixed $N$ in any case), this does not restrict the generality of our argument. For the unitary cloner we will thus consider the states obtained from Eqs. (3-4) by projecting onto a fixed number $N$ of photons in B. For the measure-and-prepare cloner we will begin by considering the simplest example, where one prepares $N$ photons in the state found by the random equatorial measurement:
\begin{equation}
\mid\Psi_{\phi}\rangle=\frac{\left(b_{\phi}^{\dagger}\right)^{N}}{\sqrt{N!}}\ket0.\label{eq:MP-phistate-equatorial}
\end{equation}
One can show that such a \MP \: cloner is asymptotically an optimal \PC \: cloner \cite{sekatskipra}.

\begin{figure}
\centering{}\includegraphics[width=\linewidth]{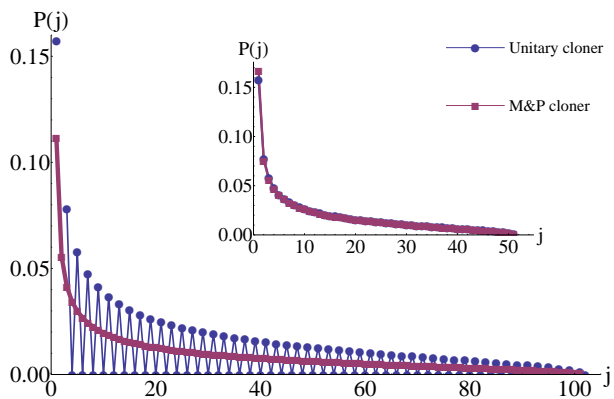}\caption{Probabilities of measuring $j$ photons in photon counter B1 of Figure 1,  for a total photon number $N=100$, for both the unitary cloner defined by Eqs. (1-4) and the measure-and-prepare cloner defined by Eqs. (\ref{eq:MP-macrostate}) and (\ref{eq:MP-phistate-equatorial}). The input state $|\theta_A,\phi_A\rangle$, which is prepared by the measurement in A, satisfies $\theta_A=\pi/2$, and the measurement in B satisfies $\theta_B=\pi/2$ and $\phi_B=\phi_A$. The probabilities for the unitary cloner have a distinctive odd-even structure, whereas those for the measure-and-prepare cloner do not. However, the inset shows that when pairs of neighboring photon numbers are put into non-overlapping bins, the two resulting probability distribution functions are almost identical.}
\label{Flo:MP_PC}%
\end{figure}

Let us begin by discussing equatorial measurements on A and B, that is $\theta_A=\theta_B=\pi/2$. Figure \ref{Flo:MP_PC} shows the \PDF \: of the outcomes $|j \rangle_{\phi_B}|N-j \rangle_{\phi_B\perp}$ for photon counting measurements in B satisfying $\Delta \phi=0$, for both the unitary cloner and the measure-and-prepare cloner described above. As we have noted previously, the unitary cloner has a distinct odd-even structure. The measure-and-prepare cloner does not. However, apart from this fine structure the two distributions are extremely similar. This is made explicit in the inset of Fig.
\ref{Flo:MP_PC}, where pairs of neighboring photon numbers were binned into non-overlapping bins. After this moderate amount of coarse-graining, the two distributions are now almost indistinguishable by the naked eye!

The results for $\Delta \phi=\pi$ look identical if one replaces $j$ by $N-j$.
Because of the phase-covariance of the two cloners, the \PDF \ for general $\Delta \phi$ can be expressed in terms of the probability distributions for $\Delta \phi=0$ and $\Delta \phi=\pi$:
\begin{align}
P(j,\Delta \phi)=&\cos ^2(\frac{\Delta\phi}{2})P(j,0)+\sin ^2(\frac{\Delta\phi}{2})P(j,\pi)\cr
+ & 2\sin(\frac{\Delta\phi}{2})\cos(\frac{\Delta\phi}{2})\sqrt{P(j,0)P(j,\pi)},
\end{align}
where $P(j,0)=\frac{j!(n-j)!}{\left(\frac{j}{2}!\frac{n-j-1}{2}!\right)^2}$ and
$P(j,\pi)=\frac{t!(n-j)!}{\left(\frac{j-1}{2}!\frac{n-j}{2}!\right)^2}$ for the unitary cloner, and $P(j,0)=\frac{2 \cdot n!}{\pi (n-j)!j!} \text{B} \left(j+1/2, n-j+3/2\right)$ and $P(j,\pi)=\frac{2 \cdot n!}{\pi (n-j)!j!} \text{B} \left(j+3/2, n-j+1/2\right)$ for the measure-and-prepare cloner, with $\text{B}(a,b)=\frac{\Gamma(a)\Gamma(b)}{\Gamma(a+b)}$ the Euler beta function.

Hence, also for general $\Delta \phi$ the probability distributions for the two cloners become essentially indistinguishable under the described binning.
This implies that in the presence of a moderate amount of coarse-graining  it is impossible to distinguish the micro-macro entangled state of Eq. (\ref{outputstate}) from a completely separable state, if the measurements for A and B are both on the equator of the Bloch sphere.

We will now discuss more general measurement directions. For general measurements, we should modify the \MP \: cloner. That is
because the \MP \: cloner defined by Eq. (6) is not a good approximation to the unitary cloner for measurements
on or near the pole of \BS . In fact, the state generated by the unitary cloner
is highly squeezed with respect to $J_z$, whereas the state generated by this particular measure-and-prepare cloner is not. This can be seen from the form of $H$ in Eq. (\ref{eq:PC_Hamiltonian}), which clearly does not change the value of $J_z$ at all, and from expanding the state of Eq. (\ref{eq:MP-phistate-equatorial}) in the $h-v$ basis. Therefore we modify the state of Eq. (\ref{eq:MP-phistate-equatorial})
by just keeping the terms with the smallest values of $J_z$ in order to have a high degree of squeezing:
\begin{align}
\ket{\Psi_{\intphi}}=\frac{1}{\sqrt{2(\tau+1)}}\sum_{k=0}^{\tau}(e^{i\phi(2k+1)} |n-k\rangle_h |n+k+1\rangle_v\nonumber\\
+e^{-i\phi(2k+1)} |n+k+1\rangle_h |n-k\rangle_v), \label{eq:G-MP-state} \end{align}
where the moduli of the coefficients of all terms can be chosen equal for small $\tau$, $n=\frac{N+1}{2}$, $\tau$ indicates how many terms we are keeping, and the state is written in the $h-v$ basis. Clearly $\tau=0$
is the most squeezed case and as it increases, the state becomes less
squeezed. The results shown below are for $\tau=2$.

We quantify the distance between the probability distributions for the two cloners by the Manhattan norm, which is defined as:
\begin{equation}
D=\underset{j}{\sum}\left|P_{u}(j)-P_{mp}(j)\right|,
\label{eq:Manhatan-distance}
\end{equation}
where $P_{u}(j)$ and $P_{mp}(j)$ are the distributions for the unitary and measure-and-prepare cloner respectively. This is a global measure of statistical difference between two probability distributions \cite{krause}. We furthermore model the \CG ing in this experiment with a basic symmetric overlapping binning, i.e. we consider coarse-grained probability distributions
\begin{equation}
\bar{P}(j)=\frac{P(j-\frac{\sigma-1}{2})+ ...+P(j)+...+P(j+\frac{\sigma-1}{2})}{\sigma},
\label{eq:CG simple model}
\end{equation}
where $\sigma$ is the bin size. For convenience we only consider odd values of $\sigma$ in the following, and we are using periodic boundary conditions for $j$. Other choices of coarse-graining lead to equivalent results.

\begin{figure}
\centering{}\includegraphics[width=\linewidth]{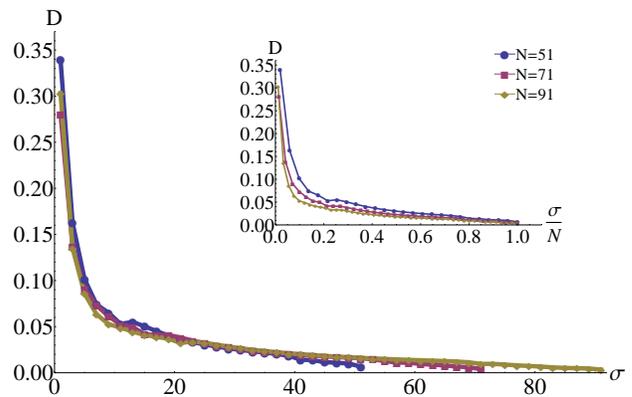}\caption{The distance between the probability distributions for the unitary and measure-and-prepare cloner, quantified by the Manhattan-norm distance $D$ of Eq. (\ref{eq:Manhatan-distance}), as a function of the bin size $\sigma$ of Eq. (\ref{eq:CG simple model}). The distance decreases with $\sigma$ in a way that is almost independent of the photon number $N$. As a consequence, as shown in the inset, the distance decreases faster and faster as a function of the relative error $\sigma/N$, when $N$ is increased. The results shown are for $\theta_A=\pi/2$, $\theta_B=\pi/12$ and $\Delta \phi=0$, but the behavior is generic, see text.
}
\label{Flo:D-sigma}%
\end{figure}

Figure \ref{Flo:D-sigma} shows
the distance $D$ between the two distributions as $\sigma$ is increased. The figure shows the results for one specific choice of angles, but we have tested over 700 different combinations, exploring all regions of phase space, and the trend is the same for all choices of angle. One sees that the distance decreases with $\sigma$ in a way that is almost independent of $N$. This means that in the present system photon number resolution at the single-photon level is essential in order to see quantum features. Note that this is consistent with the behavior for equatorial measurements discussed above (where binning of neighboring photon numbers is sufficient to make the distance unobservable). This implies that for increasing $N$ the distance decreases faster and faster in terms of the relative coarse-graining error $\sigma/N$ (see the inset of Figure \ref{Flo:D-sigma}). For increasing $N$ it thus becomes more and more difficult to distinguish the two cloners, and hence to prove micro-macro entanglement.

We have seen that a small and fixed amount of coarse-graining makes the difference between the entangled and separable micro-macro states unobservable. This is in contrast to the macro-macro entangled state discussed in the introduction, where sub-$\sqrt{N}$ resolution is sufficient to prove entanglement \cite{entlaser}. Micro-macro entanglement is thus more fragile under coarse-graining than macro-macro entanglement (at least for these examples), maybe because there is less entanglement in the state to begin with (1 ebit here compared to $\log N$ ebits for the macro-macro singlet state). From the experimental point of view this means that with current technology it is probably impossible to prove the present kind of micro-macro entanglement without supplementary assumptions for large photon numbers using photon counting measurements.

Conceptually, our results strengthen the idea that precise (non coarse-grained) measurements are generally essential for demonstrating quantum features at the mesoscopic or macroscopic level. Another well-known example is the ``cat state'', i.e. the superposition of two coherent states of a single bosonic mode \cite{Zurek-CtoQ-03}. For coherent states with large amplitudes, the ripples in the Wigner function that indicate the quantum superposition require increasingly high resolution for the quadrature measurements in order to be observable. Analyzing the necessary resolution for quadrature measurements, i.e. homodyne detection, for our present example is work for the future. However, Ref. \cite{spagnoloq} recently studied the effect of photon loss, which can be seen as a form of coarse-graining, and found results that are consistent with ours. We would also like to mention Ref. \cite{jeong}, where a Bell inequality violation is predicted for very coarse-grained measurements. However, the considered measurements use a strong nonlinear interaction that produces phase shifts of $\pi$ between subsequent photon number states, which corresponds to high resolution in a slightly more general sense. We would therefore argue that the results of Ref. \cite{jeong} are not in contradiction with the hypothesis that high resolution is essential, but rather show that the concept of coarse-graining should be refined in contexts where strong interactions are considered. A more detailed analysis of this question is also work for the future.

\begin{acknowledgments}
We thank J. Davidsen, N. Gisin, F. Sciarrino, and W. Tittel for helpful comments, and we acknowledge AITF, NSERC, \textit{i}CORE, and the Swiss NFS for financial support.
\end{acknowledgments}



\end{document}